# Estimating the Expected Influence Capacities of Nodes in Complex Networks under the Susceptible-Infectious-Recovered (SIR) Model

**Aybike ŞİMŞEK**

**Abstract**
In recent years, epidemic modeling in complex networks has found many applications, including modeling of information or gossip spread in online social networks and modeling of malware spread in communication networks. In this context, identifying the super-spreaders is an important topic. Most studies aiming to distinguish the influences of nodes use only network centrality measures and subsequently, rank the nodes accordingly. However, the dynamics of the propagation model are important in determining the influence capacity of nodes. In this study, we developed Expected Value Estimation (EVE) algorithm that estimates the expected influence of nodes under the Susceptible-Infectious-Recovered (SIR) model. The basic idea of the algorithm is that a node is more likely to influence another node over the shortest path between these two nodes than over other paths because the probability of influence exponentially decreases as the number of hops increases. Therefore, we can quickly estimate the influence capacity of the nodes by calculating the expected value using only the shortest paths and the propagation model dynamics. We compared its performance on real-world datasets with that of many well-known current centrality measures. The experiments show that the solution quality of EVE was superior to that of its competitors.

## 1. Introduction

Complex networks are highly suitable tools for modeling the real world. They have applications in many different fields such as natural sciences (Gao, Barzel, and Barabási 2016), health (Barabási, Gulbahce, and Loscalzo 2011), cyber security (Alasmary et al. 2019), economics (J. Yang et al. 2010), and social networks (S. P. Borgatti et al. 2009; Leskovec, Huttenlocher, and Kleinberg 2010; Tuğal and Karcı 2019). Moreover, epidemic modeling in complex networks has attracted attention in recent years for its many practical benefits. Based on this, the spread of a virus outbreak (such as Covid-19) can be estimated and precautions taken (Chang et al. 2021). By modeling the spread of gossip on the social media network, the spread can be prevented (Y. Yang et al. 2020; Zhang et al. 2018) and/or the preferred information made available to the maximum number of people (Banerjee, Jenamani, and Pratihar 2020). Whether one desires to minimize the spread of gossip or to maximize the spread of information, in either case, in order to do so, the set having the smallest number of the most influential individuals should be identified (Stephen P. Borgatti 2006; Kempe, Kleinberg, and Tardos 2003). The influences of these individuals should be calculated via certain epidemic models such as the Susceptible-Infectious-Recovered (SIR) in order to identify the smallest number of the most influential among them (i.e., the key players). For this, it is necessary to model the propagation by selecting each node individually as the seed. Since propagation models are stochastic, they must be repeated many times (e.g., approximately 10,000 iterations) and the average value taken. This operation requires very high processing power. On the other hand, researchers have noticed a correlation between the influence capacity of the nodes and network centrality measures, which have long been used to determine the importance of nodes in complex networks. The basic expectation here is that as a centrality measure increases, the influence capacity increases, and as the centrality measure decreases, the influence capacity decreases. Studies have turned to this approach because the calculation of centrality measures requires much less processing power than modeling the propagation thousands of times. Basic centrality measures such as Closeness, Degree, Betweenness (Freeman 1977), Katz (Katz 1953), and PageRank (Page et al. 1999) have been used and new centrality measures developed for this purpose. However, many of these measures were developed by considering only the local and global impacts of the nodes (Ghalmane, El Hassouni, et al. 2019; Lv et al. 2019; Salavati, Abdollahpouri, and Manbari 2019; Sheng et al. 2020; J. Zhao, Song, and Deng 2020) or network communities (Ghalmane, Cherifi, et al. 2019; Ghalmane, El Hassouni, et al. 2019; Salavati, Abdollahpouri, and Manbari 2019; Y. Zhao, Li, and Jin 2016; Z. Zhao et al. 2015). Recently, another approach has been adopted that combines multiple centrality measures to develop new hybrid centrality measures (Ali, Anwar, and Rizvi 2020; Alshahrani et al. 2020; Keng, Kwa, and McClain 2020; Ma et al. 2016; A. Şimşek 2021; M. Şimşek and Meyerhenke 2020; X. Wen et al. 2018; Yan, Cui, and Ni 2020). However, many of these studies ignore the dynamics of the propagation model, i.e., they only determine the importance of nodes based on the structure of the network. On the other hand, in models such as SIR, a node has a certain probability of infecting its neighbors ($\beta$). Let us consider a connected network. For small $\beta$ values, the influence of a node with a high degree of centrality is greater than for a node with a low degree of centrality because it has many neighbors. The greater the $\beta$ value, the smaller the difference between the influence of a node with lower-degree centrality and a node with higher-degree centrality because the node with lower-degree centrality may spread via its neighbors, neighbors of neighbors, and so on. The extreme point of this phenomenon is when $\beta$ is 1. In this case, although there is no change in the centrality measures of the nodes, the influence of all nodes is the same because all nodes can influence all other nodes. As a result, the influence of a node depends not only on its location on the network, but also on the dynamics of the propagation model (Liu et al. 2016).

In this study, we developed an algorithm that ranks nodes according to their influence capacity, taking into account the propagation behavior in the SIR model as well as the node's location on the network. We named our developed algorithm the Expected Value Estimation (EVE) because it is based on approximating the expected influence of each node. It is worth mentioning here that, in contrast to the centrality measures, the EVE algorithm does not calculate the importance of nodes. Instead, it calculates the approximate expected influence of the nodes via the SIR model and ranks the nodes accordingly.

Under certain epidemic models (such as SIR), it is necessary to perform intense Monte-Carlo simulations to distinguish the influence of nodes. However, if the dynamics of the SIR propagation model are taken into account, the process can be simplified by ignoring some of the behaviors of this model. Thus, as with a centrality measure, the approximate expected influence of nodes can be calculated and used to rank the nodes. In the SIR model, a node influences its neighbor nodes with the probability of beta ($\beta$). If not its direct neighbor, it is likely to influence its neighbors' neighbors with a probability of ($\beta \times \beta$). If the network is in a tree form, the probability of a node influencing another node one hop away can be calculated as $\beta^l$ since there can be only one path between each pair of nodes. Thus, the expected influence of a node can be calculated using its distance to all other reachable nodes as the sum of $\beta^l$ values. However, real networks rarely exhibit tree structures. Hence, there can be many different paths of different lengths between any two nodes. It is also very costly to use all paths to all other nodes to calculate the expected influence of a node. However, the probability of one node influencing another node decreases exponentially with the distance between them, although in practice, the value of $\beta$ is much less than 1. The natural consequence of this is $\beta^n \gg \beta^{n+1}$, where $n \in \mathbb{N}^+$. Based on this information, the expected probability of a node influencing another node can only be approximated using the shortest path between these two nodes. This is because the probability of influence calculated for routes other than the shortest path will be much lower. As with centrality measures, these calculated values can be used to distinguish the influence capacities of the nodes.

The Dynamics-Sensitive (DS) centrality is a similar approach that combines network structure and epidemic model dynamics for ranking nodes (Liu et al. 2016). The DS considers all possible random walks between two nodes for estimating the infectious probabilities of nodes. This method may work well for small β values. However, for larger β values, it will overestimate the infection capabilities of nodes. Another study using Multi-Dimensional Social Influence (MSI) reported that the application of centrality measures only was not sufficient (Zhuang, Li, and Zhuang 2021). The MSI combines structure-based (centrality measures), information-based (popularity of the information, the type of information, etc.), and action-based (interactive frequency, correlation of neighbors, etc.) factors to identify influential users in online social networks. The MSI outperforms its competitors in terms of ranking accuracy. It should be noted that MSI focuses on online social networks rather than general complex networks. One study considers the Diffusion Centrality (DC) model (Kang et al. 2016), which measures how far a node spreads a particular property under a given diffusion model. Similar to MSI, DC is a topic-aware approach.

In summary, centrality-based methods ignore the propagation model dynamics. The studies that consider diffusion model dynamics such as MSI and DC are topic-dependent. Although DS works on general complex networks, it overestimates at high β values. To this end, the main contributions of the EVE algorithm are as follows:

- EVE takes into account the SIR dynamics as well as the node's location on the network (by using the shortest paths between the nodes).
- While the centrality measures indirectly distinguish the nodes according to their influence capacity; EVE directly measures the approximate influence capacity of nodes under SIR.
- EVE is fast and easy to implement. The main time-consuming feature of the algorithm is the calculation of the shortest paths.
- Experiments on several real and synthetic datasets showed that EVE ranks nodes by their influence capacities, and can detect influential users.

The known limitations of EVE are as follows:

- It is not a general centrality measure. It depends on the SIR model.
- EVE may underestimate the influence capacities of nodes on high-density networks because there may be many paths between any two nodes.

## 2. Preliminaries

Before discussing the details of EVE, it would be useful to give some preliminary information.

Let $G = (V, E)$ be an undirected, unweighted graph (network). Here, V is the set of nodes (vertices), and E is the set of edges (links).

**Definition 1. Susceptible-Infectious-Recovered Model**: The Susceptible-Infectious-Recovered (SIR) model is a well-known model used for population-based epidemic modeling. In recent years, due to their popularity, SIR and SIR variations have been applied to network topologies (Tolić, Kleineberg, and Antulov-Fantulin 2018). In the SIR model, nodes are found in one of three states: Susceptible, Infected, or Recovered. The transition of nodes between states occurs according to certain probabilities. With a probability of β, susceptible nodes are more likely to be infected by neighbors who are already infected. Infected nodes are also likely to go into a recovered state with a probability of γ. Initially, all other nodes are in a susceptible state, except for nodes that carry the disease (i.e., those that are infected). Starting from the nodes that are initially infected (called ‘seed nodes’), the disease spreads over the network. After a certain period of time, there are no remaining infected nodes on the network and thus, the model is terminated.

**Definition 2 Kendall's tau Ranking Correlation Coefficient** (Kendall 1938): Let $(a_i, b_i)$ and $(a_j, b_j)$ be tuples of joint A and B ranking lists. If $a_i > a_j$ and $b_i > b_j$ or $a_i < a_j$ and $b_i < b_j$ , then the tuples are concordant. If $a_i > a_j$ and $b_i < b_j$ or $a_i < a_j$ and $b_i > b_j$ , then the tuples are discordant. If $a_i = a_j$ or $b_i = b_j$ , then the tuples are neither concordant nor discordant. Tau is defined in (1).

$$\mathrm{tau} = \frac{N_c - N_d}{0.5N(N-1)} \tag{1}$$

Here, $N_c$ is the number of concordant pairs, $N_d$ is the number of discordant pairs, and N is the number of all combinations. Positive tau values indicate a positive correlation, and negative tau values indicate a negative correlation.

**Definition 3 Ranking Monotonicity** (Bae and Kim 2014): Monotony is a metric indicating how well the centrality measure assigns each node to different rank levels. The ranking monotonicity (RM) will be '1' if all nodes are assigned to a different ranking level. If all nodes are assigned to the same ranking level, the RM will be '0'. Of course, for a centrality measure, the closer it is to RM 1, the better. The RM is calculated as in (2):

$$RM(L) = \left(1 - \frac{\sum_{r \in L} n_r(n_r - 1)}{n(n-1)}\right)^2 \tag{2}$$

Here, n is the length of the L-ranking list and $n_r$ the number of elements assigned to the same r rank.

## 3. EVE

The working principle of EVE is based on expected value calculation. Therefore, it is useful to first look into the details of how a node infects its neighbor nodes in SIR and how this node recovers. This situation is shown for one iteration in Algorithm 1 (Rossetti et al. 2018). The node u in the algorithm was initially selected as the infected node or one infected at any point in time.

**Algorithm 1. Infection and Recover States of SIR**

1 sn = susceptible neighbors of node u
2 **for each** v **in** sn
3 rnd = random number in [0.0, 1.0]
4 **if** rnd < $\beta$ **then**
5 mark v as infected
6 **end for**
7 rnd = random number in [0.0, 1.0]
8 **if** rnd < $\gamma$ **then**
9 mark u as recovered

According to Algorithm 1, node u infects its neighbors with a probability of β. After node u has infected its neighbors, this node recovers with a probability of γ. If γ = 1, node u has absolutely only one attempt to infect its neighbors since it will not be in the infected state in the next iteration. If γ = 0.5, roughly speaking, node u has two attempts to infect its neighbors with a probability of 0.5 since it will be in the infected state in the next iteration. If we generalize, node u has at least $1/\gamma$ attempts to infect its neighbors. Since the probability of node u infecting its neighbors is β, the expected value of infecting a neighbor by node u would be $1/\gamma$ times β, i.e., $\beta/\gamma$.

Let us explain the situation in Figure 1, where different topologies are shown. Notice that Figure 1-a, b, and c are tree structures. Therefore, there is only one path between all nodes.

In Figure 1-a, let node u initially be selected as a seed (infected). The expected influence value (ev) of node u becomes $ev(u) = 1 + \beta/\gamma$. Here, 1 has been added as node u is already infected.

Figure 1-b shows the expected influence value (ev) of the node as ev(u) = $1 + \beta/\gamma +$ (probability of u infecting y). In order to infect node y, node u must infect node x. Next, node x must infect node y. The probability of these two events happening together can be obtained by multiplying the probabilities of their respective occurrence. Thus, the expected value of u infecting node y is $\left(\beta/\gamma \times \beta/\gamma\right)$, i.e., $\left(\beta/\gamma\right)^2$. Hence, the expected influence value (ev) of node u becomes ev(u) = $1 + \beta/\gamma + \left(\beta/\gamma\right)^2$. For Figure 1-c, the expected influence value (ev) of node u is ev(u) = $1 + 2 \times \left(\beta/\gamma\right) + 2 \times \left(\beta/\gamma\right)^2$.

The expected value of a node infecting another node decreases exponentially with the distance between them. If we generalize the ev calculation, we get (3).

$$\mathrm{ev(u)} = 1 + \mathrm{nn}_1 \times \left(\beta/\gamma\right) + \mathrm{nn}_2 \times \left(\beta/\gamma\right)^2 + \cdots + \mathrm{nn}_\mathrm{h} \times \left(\beta/\gamma\right)^\mathrm{h} \tag{3}$$

Here, nn is the size of the set of node u's neighbors at h-hop distance. The situation is a little different in Figure 1-d. Node y is both a 1-hop and a 2-hop neighbor of node u. Therefore, node u can infect node y directly, as well as through node x. Thus, the expected value of node u infecting node y is the sum of these two possibilities, or 1 at most. Ultimately, the expected influence of node u becomes

$\mathrm{ev(u)} = 1 + \left(\beta/\gamma\right) + \max\left(1, \left(\left(\beta/\gamma\right) + \left(\beta/\gamma\right)^2\right)\right)$. Here, max() function returns the largest of its parameters. Let us explain why we use the max function here. For example, if $\beta/\gamma = 1$, the expected value of node u infecting node y would be 2. However, this value can be at most 1, since once a node is infected, it cannot be infected again.

In large and complex networks, there can be many different paths having different lengths from one node to another. As stated in the Introduction, in practice, the value of β is much less than 1. Therefore, the probability of a node influencing another node over a longer path is much lower than the probability of influencing that node over a shorter path. Consequently, it is quite costly to consider all paths. Instead, only the shortest paths can be considered to increase the computation speed. Thus, as in Figure 1-e, the (x, y) edge is ignored and the approximate ev can be calculated using (3). However, instead of changing the structure of the graph, only neighbors with h-shortest path-hop distance can be included when creating $\mathrm{nn}_\mathrm{h}$sets. Thus, it is guaranteed that $\mathrm{nn}_\mathrm{a} \cap \mathrm{nn}_\mathrm{b} = \emptyset$. Here, $\mathrm{a} \neq \mathrm{b}$ and $\mathrm{a, b} \in \{1 \ldots \mathrm{h}\}$. If $\mathrm{spn}_\mathrm{h}$ are the sets created by selecting only neighbors with h-shortest path-hop distance, we can calculate the measure we call EVE as in (4).

$$\mathrm{EVE(u)} = 1 + \mathrm{spn}_1 \times \left(\beta/\gamma\right) + \mathrm{spn}_2 \times \left(\beta/\gamma\right)^2 + \cdots + \mathrm{spn}_\mathrm{h} \times \left(\beta/\gamma\right)^\mathrm{h} \tag{4}$$

In (4), paths other than the shortest paths are not taken into account. In the literature, β is usually taken as very small (e.g., ≤0.1) and γ as large (e.g., = 1). The corollary of this is $\left({\beta}/{\gamma}\right)^{l} \gg \left({\beta}/{\gamma}\right)^{l+1}$, where $l \in \mathbb{N}^{+}$. Thus, it can be considered reasonable to ignore paths other than the shortest paths.

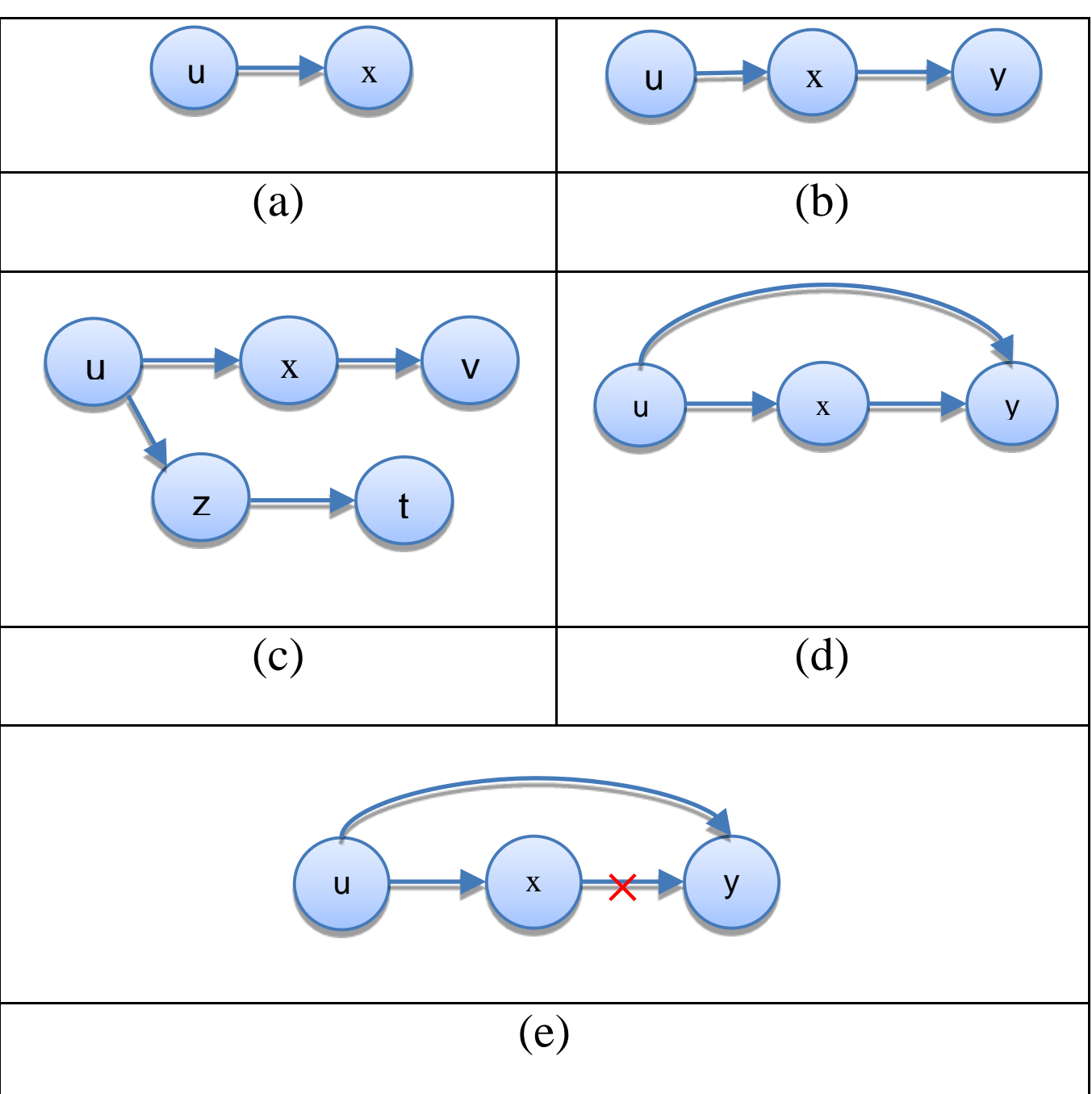

Fig. 1. EVE calculation on particular paths: (a), (b), (c) every node belongs to only one h-hop neighborhood, (d), (e) node y belongs to different h-hop neighborhoods.

In practice, EVE can be calculated as in Algorithm 2.

**Algorithm 2. EVE**

1 **FunctionEVE**(G: Graph, β, γ)
2 **Begin**
3 L = { } // L is a (key, value) dictionary as L[node] =EVE
4 SP = dictionary of all pairs shortest path of G.
5 //SP is a dictionary [source, destination] =length.
6 //If there is at least one path between two nodes then //SP[node, node] is a number. Otherwise, it is ∞.
7 V = G's set of nodes
8 **for each** u **in** V
9 EVE = 0
10 **for each** v **in** V
11 **if** SP[u,v]≠ ∞ **then**
12 EVE = EVE + **Power**($^{\beta}/_{\gamma}$, SP[u,v])
13 L[u] =EVE
14 **Sort** L descending order by value
15 **Return** key list of L
16 **End**

The Sort function in Algorithm 2 sorts the dictionary entries according to their values in descending order. The Power function takes two parameters such as x and y and returns the value $x^y$. As a result, Function EVE returns the list of nodes sorted in descending order according to their EVE values.

Figure 2 demonstrates how Algorithm 2 works for a node. The shortest paths from node u to all other nodes are shown with bold edges. The pale gray edges are the back edges. The values written on top of the nodes except node u show how much these nodes contribute to the EVE value of node u. The sum of these values gives the EVE value of node u. Algorithm 2 does this for each node in the network and calculates the EVE for all nodes.

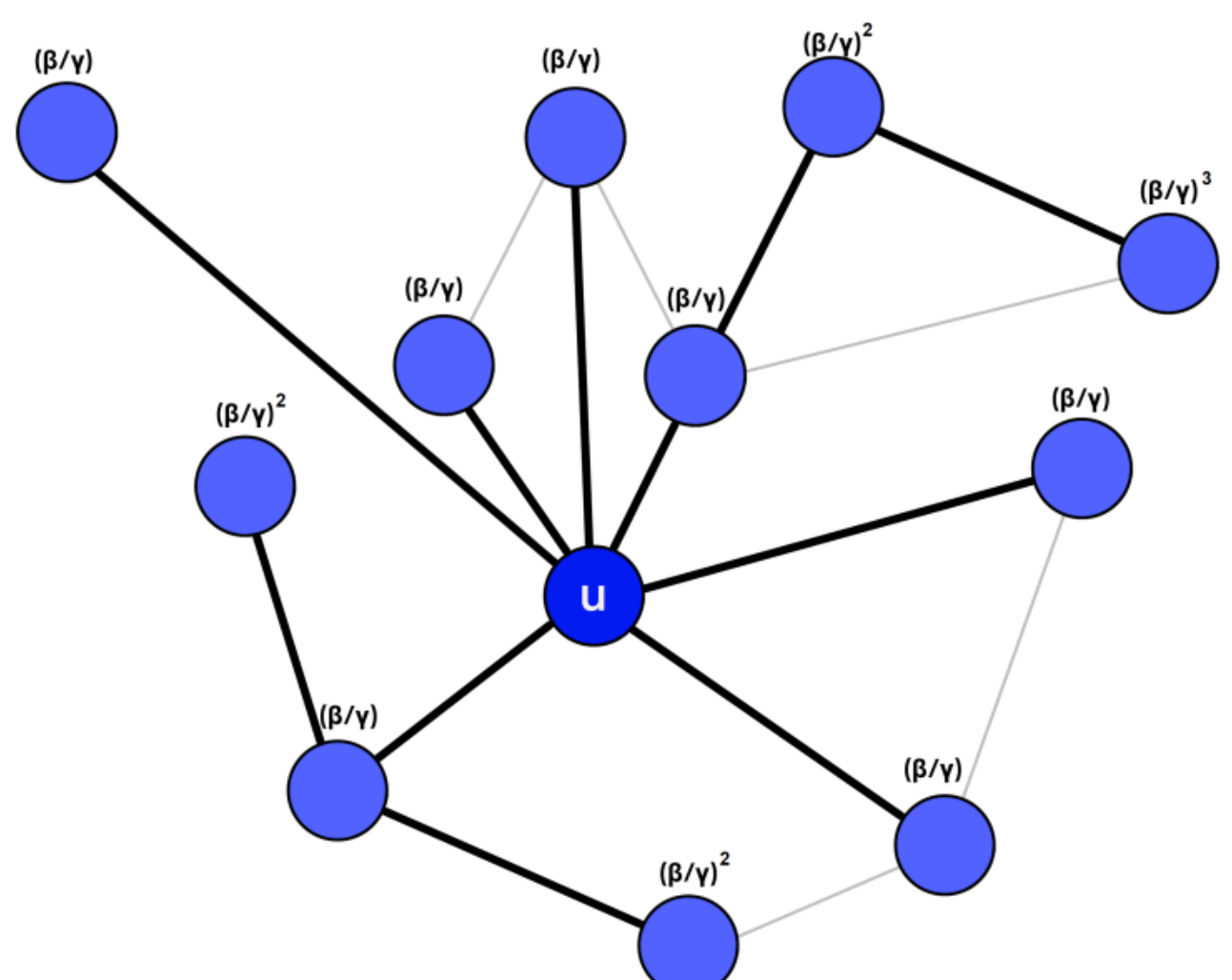

Fig. 2. A case of calculating EVE for a node on a sample graph.

## 4. Experiments

To evaluate the performance of EVE, we determined six competitor centrality measures and experimented with different SIR settings over one synthetic and eight real-world datasets. First, let us look at the competing centrality measures and datasets.

### 4.1. Centrality measures

*Degree Centrality (DC)* is the ratio of the degree of the node to the number of nodes in the graph minus one (Newman 2018).

*Eigenvector Centrality (EC*) is used to determine the importance of a node in the network. The idea behind the EC is that the more adjacent a node is to the important nodes, the more important it is (Bonacich 1987).

*Closeness Centrality (CC)* is a measure of how close a node is to other nodes (Sabidussi 1966). The closer the node is to other nodes, the larger the CC.

*Betweenness Centrality (BC)* is the proportional information on how many of the shortest paths between all pairs are through a node (Freeman 1977).

*Gravitational Centrality (GC)* is a centrality measure inspired by Newton's gravitational formula (Ma et al. 2016). Instead of the mass in the original formula, it uses the k-shell values of the nodes and instead of the distance, it uses the length of the shortest path between nodes. Its formula is as in (5).

$$GC_i = \frac{ks_i \times ks_j}{\sum_{j \in N} d(j,i)} \quad (5)$$

Here, $d(\cdot)$ is the length of the shortest path between nodes i and j and N is the set of 3-hop neighbors of node i.

*Multi-local dimension (MLD)* is a state-of-the-art centrality measure proposed by Wen et al. (T. Wen, Pelusi, and Deng 2020) that considers a node as the center and calculates the ratio of this node's neighbors up to the r-hop distance to the number of all nodes for different radius (r) values. It then calculates a centrality measure for the node based on this value. For the details of MLD, the related study can be examined.

### 4.2. Datasets

We used one synthetic (Barabasi-Albert, BA) and eight real-world networks for the experiments. The properties of the networks are given in Table 1. All the real-world datasets are taken from http://networkrepository.com (Rossi and Ahmed 2015).

Table 1. Network dataset features

| Dataset | $\|V\|$ | $\|E\|$ | $\langle K \rangle$ | $K_{max}$ | Density |
|---|---|---|---|---|---|
| BA | 1000 | 9900 | 19.8 | 198 | 0.020 |
| Ca-GrQc | 5242 | 14496 | 5.53 | 81 | 0.0010544 |
| Email-Enron | 143 | 623 | 8 | 42 | 0.0613612 |
| Email-Univ | 1133 | 5451 | 9.62 | 71 | 0.0085002 |
| inf-power | 4941 | 6594 | 2.66 | 19 | 0.000540303 |
| inf-USAir97 | 332 | 2126 | 12.80 | 139 | 0.0386925 |
| rt_alwefaq | 4171 | 7123 | 3.41 | 879 | 0.000818602 |
| rt_bahrain | 4676 | 8007 | 3.42 | 261 | 0.000732378 |
| rt_damascus | 3052 | 3881 | 2.54 | 648 | 0.000833579 |

### 4.3. Performance Comparison with Centrality Measures

We evaluated the performance of EVE and the competitor centrality measures from different angles. First, we looked at the Kendall ranking performances. We then compared their Monotonicity performances. Finally, we looked at how many of the nodes in the top 5% of the ranking lists created by the centrality measures corresponded to the ranking lists created according to the SIR simulations.

We applied the SIR model to measure the influences of the nodes. We set $\gamma = 1$, and we tried different values for $\beta$ around the epidemic threshold ($\beta_{th}$). The epidemic threshold is calculated as in (6) (Li et al. 2019).

$$\beta_{th} \approx \frac{\langle k \rangle}{\langle k^2 \rangle - \langle k \rangle} \quad (6)$$

Here, $\langle k \rangle$ denotes the average degree, and $\langle k2 \rangle$ denotes the second-order moment of the degree distribution (Li et al. 2019).

In the SIR simulations, we set each node as the only infected node in the network. We ended the simulations when there were no infected nodes left in the network. At the end of each simulation, we took the number of recovered nodes in the network as the influence of the node selected as the single infected node at the beginning of that simulation. We repeated the simulation for each node 1000 times and took the average value of their influences as the final SIR score. For the simulations we used Python and NetworkX (Hagberg, Schult, and Swart 2008).

**Kendall ranking**: the ranking performances of EVE and the competitor centrality measures for $\beta = \beta_{th}$ are shown in Table 2. The best results are emphasized in bold. Ranking performances were calculated using Definition 2, as the Kendall's tau ranking correlation coefficient. The ranking list created by the centrality measures and the list created by SIR simulations were used in the calculations.

The best results were given by EVE in six experiments, by GC in two experiments, and by EC in one experiment. In addition, the EVE tau values in all experiments were very close to 0.8 or higher. The more detailed results are shown in Figure 3.

Table 2. Kendall's tau correlation coefficient results of the centrality measures for $\beta = \beta_{th}$.

| **Dataset** | **DC** | **EC** | **CC** | **BC** | **GC** | **MLD** | **EVE** |
|---|---|---|---|---|---|---|---|
| BA | 0.52 | **0.56** | 0.55 | 0.49 | 0.56 | 0.18 | 0.55 |
| Ca-GrQc | 0.68 | 0.53 | 0.50 | 0.37 | 0.73 | 0.48 | **0.74** |
| Email-Enron | 0.75 | 0.79 | 0.66 | 0.45 | **0.81** | 0.64 | 0.77 |
| Email-Univ | 0.72 | 0.77 | 0.73 | 0.59 | **0.77** | 0.66 | 0.76 |
| inf-power | 0.42 | 0.45 | 0.26 | 0.29 | 0.55 | 0.48 | **0.56** |
| inf-USAir97 | 0.73 | 0.75 | 0.69 | 0.56 | 0.76 | 0.65 | **0.76** |
| rt_alwefaq | 0.23 | -0.05 | -0.05 | 0.29 | 0.38 | 0.0 | **0.39** |
| rt_bahrain | 0.41 | -0.19 | -0.18 | 0.39 | 0.63 | 0.0 | **0.65** |
| rt_damascus | 0.38 | -0.12 | -0.12 | 0.30 | 0.50 | 0.0 | **0.51** |

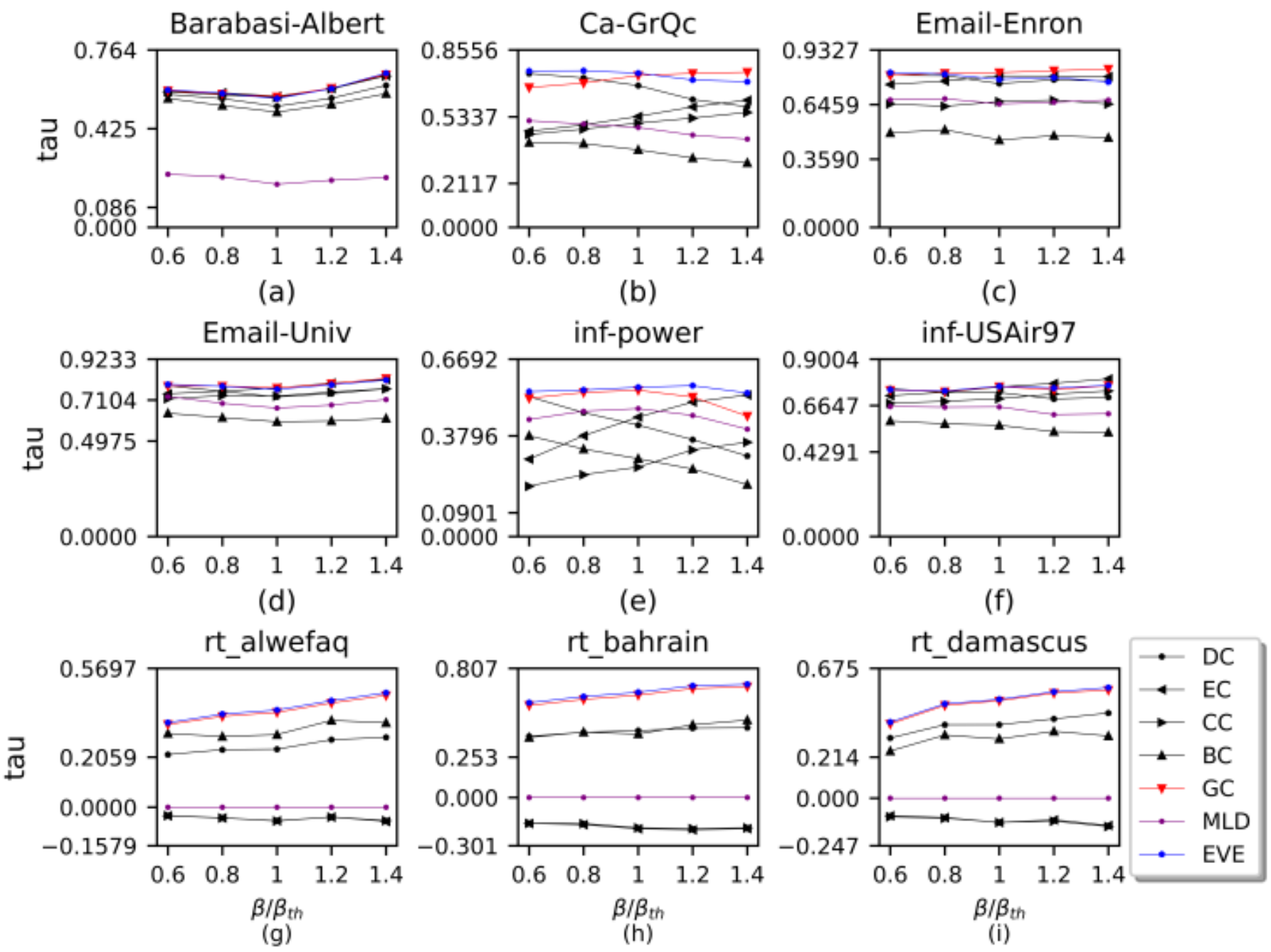

Fig. 3. Kendall's tau correlation coefficient results of the centrality measures.

**Ranking Monotonicity**: the monotonicity values of the ranking lists created by EVE and the competitor centrality measures are shown in Table 3. The values were calculated using Definition 3. Since the ranking lists created by the centrality measures depend only on the network structure, their monotonicity values were calculated only once for each dataset. The ranking list created by EVE is dependent on β. Therefore, its monotony values should be calculated for each β value. On the other hand, the monotonicity of the EVE at different β values is very close to the monotonicity of the EVE at $\beta = \beta_{th}$. For the sake of brevity, we only give the monotonicity of the EVE values at $\beta = \beta_{th}$. The monotonicity values calculated for EVE were 1 in three experiments and very close to 1 in the other three experiments. Moreover, the EC, CC, GC, and MLD also yielded successful results. In the retweet networks (rt_alwefaq, rt_bahrain, rt_damascus) there are hub nodes with a very high degree compared to other nodes. Many other nodes are linked only to these hubs. Therefore, there is only one path between most of the nodes. These networks resemble a tree structure, and many paths they have are naturally the shortest paths. On the other hand, GC and EVE are calculated using the shortest paths. We think that GC and EVE have given competitive results for the retweet networks for these two reasons.

Table 3. Monotonicity values of the centrality measures.

| Dataset | DC | EC | CC | BC | GC | MLD | EVE |
|---|---|---|---|---|---|---|---|
| BA | 0.92 | 1.0 | 0.99 | 1.0 | 1.0 | 1.0 | 1.0 |
| Ca-GrQc | 0.96 | 0.99 | 1.0 | 0.78 | 1.0 | 1.0 | 1.0 |
| Email-Enron | 0.99 | 1.0 | 1.0 | 0.99 | 1.0 | 1.0 | 1.0 |
| Email-Univ | 0.97 | 0.99 | 0.99 | 0.99 | 0.99 | 0.99 | 0.99 |
| inf-power | 0.80 | 0.68 | 0.99 | 0.95 | 0.99 | 0.99 | 0.99 |
| inf-USAir97 | 0.98 | 0.99 | 0.99 | 0.94 | 0.99 | 0.99 | 0.99 |
| rt_alwefaq | 0.69 | 0.96 | 0.99 | 0.08 | 0.26 | 0.06 | 0.26 |
| rt_bahrain | 0.76 | 0.88 | 0.99 | 0.15 | 0.48 | 0.19 | 0.48 |
| rt_damascus | 0.39 | 0.31 | 0.95 | 0.05 | 0.23 | 0.09 | 0.23 |

Finally, we examined how many of the nodes in the top x% of the ranking lists created by the centrality measures coincided with the nodes in the top x% of the ranking list created according to the SIR simulations. The results are shown in Tables 4-5. The best results are emphasized in bold. Nodes in the top-rank levels formed by the centrality measures are expected to be the more influential nodes. Therefore, the nodes at the top of the list and those at the top of the ranking list created according to the SIR simulations should be the same. According to the results, EVE outperformed the competitors in four experiments for the top 3% and top 5% of the ranking lists.

Table 4. Number of matching nodes in the top 3% of the ranking list created according to SIR simulations with THOSE created by the centrality measures.

| Dataset | DC | EC | CC | BC | GC | MLD | EVE |
|---|---|---|---|---|---|---|---|
| BA | 24 | 25 | **26** | 24 | 25 | 23 | 24 |
| Ca-GrQc | 18 | **19** | 13 | 2 | 21 | 0 | **19** |
| Email-Enron | 1 | **4** | 1 | 1 | 3 | 2 | 1 |
| Email-Univ | 20 | **24** | 17 | 14 | 19 | 16 | 21 |
| inf-power | 47 | **88** | 16 | 13 | 86 | 48 | 50 |
| inf-USAir97 | **8** | **8** | 6 | 4 | **8** | 7 | **8** |
| rt_alwefaq | 57 | 17 | 12 | 42 | 49 | 29 | **61** |
| rt_bahrain | 106 | 11 | 5 | 58 | 90 | 23 | **110** |
| rt_damascus | 34 | 3 | 4 | 26 | **36** | 26 | **36** |

Table 5. Number of matching nodes in the top 5% of the ranking list created according to SIR simulations with THOSE created by the centrality measures.

| Dataset | DC | EC | CC | BC | GC | MLD | EVE |
|---|---|---|---|---|---|---|---|
| BA | 40 | **41** | **41** | **41** | **41** | 39 | **41** |
| Ca-GrQc | 36 | 26 | 20 | 7 | 38 | 0 | **37** |
| Email-Enron | 4 | 4 | 3 | 3 | 4 | **5** | 3 |
| Email-Univ | 37 | 36 | 35 | 31 | **38** | 34 | 36 |
| inf-power | 80 | **150** | 46 | 39 | 134 | 91 | 101 |
| inf-USAir97 | 12 | 12 | 10 | 9 | **13** | 12 | 12 |
| rt_alwefaq | 77 | 53 | 43 | 55 | 68 | 71 | **82** |
| rt_bahrain | 161 | 18 | 6 | 97 | 158 | 63 | **181** |
| rt_damascus | 86 | 24 | 23 | 49 | 83 | 63 | **93** |

## 5. Discussion and Conclusions

In this study, we proposed a novel approach that approximates the influences of nodes in complex networks under the SIR propagation model using the shortest paths between nodes and then applied this to rank the nodes. As a result of nine datasets and five different SIR settings, EVE performed better than well-known and state-of-the-art centrality measures. EVE demonstrated that the expected influences of nodes could be better distinguished by using the parameters of the propagation model and the shortest paths (without using the centrality measures of the nodes).

If we analyse EVE's algorithm, we see that its algorithmic complexity as $O(|V|^3)$. The cost of calculating all pairs shortest paths with a well-known algorithm such as Floyd's algorithm is $|V|^3$. The nested for loops add $|V|^2$cost. So, in the worst case, EVE's algorithmic complexity will be $O(|V|^3)$.

EVE is calculated using the shortest paths between nodes just like GC. This means that all other paths are ignored. In dense networks, there can be many different paths other than the shortest path between two nodes. Therefore, ignoring these paths increases the difference (error) between EVE and the actual expected influence.